\documentclass[aps,prd,10pt,a4paper,superscriptaddress,preprintnumbers,showpacs,notitlepage,eqsecnum,nofootinbib]{revtex4-1}

\usepackage[english]{babel}
\usepackage{graphicx}
\usepackage{amsmath,amssymb}
\usepackage{xcolor}
\usepackage{eufrak}
\usepackage{caption}
\usepackage{subcaption}
\usepackage{hyperref}    
\usepackage{stmaryrd}

\def\sqr#1#2{{\vcenter{\vbox{\hrule height.#2pt\hbox{\vrule
width.#2pt height#1pt \kern#1pt\vrule width.#2pt}\hrule height.#2pt}}}}

\makeatletter

\@addtoreset{equation}{section}
\makeatother
\begin{document}

\title{On the motion of hairy black holes\\
in Einstein-Maxwell-dilaton theories}
\author{F\'elix-Louis Juli\'e}

\affiliation{APC, Universit\'e Paris Diderot,\\
CNRS, CEA, Observatoire de Paris, Sorbonne Paris Cit\'e\\
 10, rue Alice Domon et L\'eonie Duquet, F-75205 Paris CEDEX 13, France.}

\date{November 29th, 2017}

\begin{abstract}
Starting from the static, spherically symmetric black hole solutions in massless Einstein-Maxwell-dilaton (EMD) theories, we build a ``skeleton" action, that is, we phenomenologically replace black holes by an appropriate effective point particle action, which is well suited to the formal treatment of the many-body problem in EMD theories. We find that, depending crucially on the value of their scalar cosmological environment, black holes can undergo steep ``scalarization" transitions, inducing large deviations to the general relativistic two-body dynamics, as shown, for example, when computing the first post-Keplerian Lagrangian of EMD theories.
\end{abstract}

\maketitle

\section{Introduction}

The observations of gravitational waves emitted by the coalescence of binary black holes by the LIGO-Virgo collaboration \cite{Abbott:2016blz,Abbott:2016nmj,Abbott:2017vtc,Abbott:2017oio}, together with the recent joint detection of the coalsecence of a binary neutron star system with electromagnetic counterparts \cite{TheLIGOScientific:2017qsa}, have opened a new era in gravitational wave astronomy. In particular, these detections provide the opportunity to challenge general relativity (GR), as well as modified gravities (i.e., gravity theories beyond GR), in the strong field regime of the coalescence of two compact objects, in diversified configurations, and at various redshifts.\\

In GR, post-Newtonian (PN) expansions of Einstein's equations (in the small orbital velocity, weak field limit) are suitable to describe analytically the inspiral phase of binary systems and associated gravitational waveforms. They often rely on the very convenient ``skeletonization" of compact bodies, that is, on their phenomenological reduction to effective point particles, that follow the geodesics of the metric they produce (although necessitating regularization schemes due to the artificial introduction of Dirac distributions in Einstein's equations), see, e.g., \cite{lesHouches} and \cite{Blanchet:2013haa}.\\

The generalization of skeletonization to modified gravities was introduced by Eardley in the simplest case of (massless) scalar-tensor theories in \cite{Eardley}. There, he proposed to endow each point particle with and effective ``mass function", $m_A(\varphi)$, that depends on the (regularized) value of the scalar field evaluated at its location, which in turn identifies to the scalar environment of the skeletonized body. The resultant simplification of the formal treatment of the two-body problem enabled to address it up to 2.5 PN order \cite{Damour:1992we, Damour:1995kt, Mirshekari:2013vb}, or, adopting the terminology of \cite{Damour:1992we}, 2.5 post-Keplerian (PK) order, to highlight the fact that (strong) self-gravity effects are encompassed in the mass function $m_A(\varphi)$.
The explicit computation of $m_A(\varphi)$ for a given body is well-known in the case of neutron stars. It is based on the (numerical) integration of the field equations inside the star, depending on its internal structure (e.g., its equation of state), up to the exterior of the star, where the fields are matched to the near-worldine fields sourced by the effective point particle, see, e.g., \cite{ Damour:1992hw, Barausse:2012da, Palenzuela:2013hsa}; see also \cite{Julie:2017ucp}. 
On the other hand, in that class of theories, its is known that black holes cannot carry scalar ``hair", and hence, are reduced to Schwarzschild's \cite{Hawking:1972qk}.\\

In the present paper, we propose to go beyond what was done so far in scalar-tensor theories (ST), and consider the class of Einstein-Maxwell-dilaton theories (EMD). They consist in supplementing ST theories with a (massless) vector gauge field that is non-minimally coupled to the scalar field. Note that this vector does not necessarily correspond to the Maxwell field of electrodynamics, but must rather be considered as a new ``graviphotonic" degree of freedom of gravity.
These EMD theories will be at the center of our attention, since, contrarily to ST theories, they predict the existence of black holes that differ from the general relativistic ones through the presence of vector and induced scalar ``hairs" \cite{Gibbons:1987ps, Garfinkle:1990qj, Frolov:1987rj, Horne:1992zy}. Our aim will therefore be to address the problem of motion of two compact bodies, including these ``hairy" black holes, by generalizing the Eardley-type point particle action to EMD theories.\\

The dynamics of binary black holes in EMD theories has been studied numerically in \cite{Hirschmann:2017psw}. There, the authors found that for very small values of their scalar cosmological environment $\varphi_0$, the influence of the scalar interaction can be largely neglected in understanding their dynamics, which is therefore hardly distinguishable from its general relativistic counterpart. Now, although these conclusions will coincide with the (analytical) results obtained in the present paper, we shall hint that, when $\varphi_0$ is increased (which, as we shall argue, is physically reasonnable), black holes can undergo steep ``scalarization" transitions, i.e. become strongly coupled to the scalar and vector fields, inducing large deviations to the general relativistic two-body dynamics.\\

The paper is organized as follows: in section \ref{sectionLagrangian} we introduce EMD theories and build an appropriate point particle action, endowing them with a mass function $m_A(\varphi)$ (``\`a la" Eardley) and a constant charge $q_A$. We then compute the resultant two-body Lagrangian at post-Keplerian order (1PK), which describes the conservative part of the dynamics of arbitrary compact bodies in interaction. In section \ref{sectionSkeletonization}, we present a class of EMD black hole solutions with an ``electric" charge and dilatonic ``hair". By means of an appropriate matching condition, we then show how to skeletonize these black holes, that is, how to compute explicitely the corresponding mass function and charge. Finally, in section \ref{section_Sensitivities}, we specify the values of the post-Keplerian couplings that enter the two-body Lagrangian for the black holes mentioned above, and highlight that, depending on their cosmological scalar environment $\varphi_0$, black holes can transition between two extremal regimes: a ``Schwarzschild"-like behaviour, which makes them undistinguishable from their general relativistic counterparts, and a ``scalarized" regime, where their couplings to the scalar and vector fields reach the order of unity, inducing large deviations from general relativity.

\section{The two-body problem in Einstein-Maxwell-dilaton theories\label{sectionLagrangian}}

\subsection{The Einstein-Maxwell-dilaton skeleton action\label{subsectionSkeleton}}
The Einstein-Maxwell-dilaton (EMD) theories describe general relativity supplemented by a massless scalar field and a massless vector gauge field. In vacuum, the EMD action is conveniently written in the Einstein frame as (setting $G\equiv c\equiv 1$)
\begin{equation}
S_{\rm vac}[g_{\mu\nu},A_\mu,\varphi]=\frac{1}{16\pi}\int d^4x\sqrt{-g}\,\bigg(R-2g^{\mu\nu}\partial_\mu\varphi\partial_\nu\varphi-e^{-2a\varphi}F^{\mu\nu}F_{\mu\nu}\bigg)\ ,\label{EMDvacuumAction}
\end{equation}
where $R$ is the Ricci scalar, $g=det\,g_{\mu\nu}$, and where $F_{\mu\nu}=\partial_\mu A_\nu-\partial_\nu A_\mu$. \\

 The fundamental parameter $a$ induces a coupling between the vector and scalar fields.\footnote{Note that $a=0$ reduces to Einstein-Maxwell theory minimally coupled to a scalar field, while $a\neq 0$ is motivated by low-energy limits of string theory, see, e.g., \cite{Garfinkle:1990qj}; $a=\sqrt{3}$ is equivalent to the dimensional reduction of the Kaluza-Klein theory \cite{Frolov:1987rj} \cite{Horne:1992zy}.\\}
 In consequence, vacuum EMD theories encompass the usual $U(1)$ gauge symmetry,
$A_\mu\rightarrow A_\mu+\partial_\mu\chi$,
$\chi$ being an arbitrary scalar function, while breaking the scalar ``shift symmetry", $\varphi\rightarrow\varphi+cst$. As we shall see in the following, this will be crucial in endowing EMD black holes, that is, vacuum solutions of the theory, with scalar hair.\\ \\
In this paper, we shall restrict ourselves to $a>0$ only, since (\ref{EMDvacuumAction}) is symmetric under the simultaneous redefinitions $a\rightarrow -a$ and $\varphi\rightarrow -\varphi$.\\
 
 The vacuum field equations follow from the variation of (\ref{EMDvacuumAction}),
 \begin{subequations}
 \begin{align}
&R_{\mu\nu}=2\partial_\mu\varphi\partial_\nu\varphi+2e^{-2a\varphi}\left(F_{\mu\alpha}F_\nu^{\ \alpha}
-\frac{1}{4}g_{\mu\nu}F^2\right)\ ,\\
&\nabla_\nu\left(e^{-2a\varphi}F^{\mu\nu}\right)=0\ ,\label{maxwellVacuum}\\
&\Box\,\varphi=-\frac{a}{2}e^{-2a\varphi}F^2\ ,
\end{align}\label{vacuumEqs}
\end{subequations}
where $F^2\equiv F_{\mu\nu}F^{\mu\nu}$, where $\nabla_\mu$ denotes the covariant derivative associated to $g_{\mu\nu}$ and where $\Box\equiv\nabla_\mu\nabla^\mu$.\\

In presence of matter, the Einstein-frame action becomes
\begin{equation}
S=S_{\rm vac}+S_{\rm m}[\Psi,\mathcal A^2(\varphi)g_{\mu\nu},A_\mu]\ ,\label{actionEMD}
\end{equation}
where $\Psi$ generically denotes matter fields, that are minimally coupled to the gauge vector $A_\mu$ and to the Jordan metric, $\tilde g_{\mu\nu}=\mathcal A^2(\varphi)g_{\mu\nu}$, $\mathcal A(\varphi)$ being a scalar function of $\varphi$ that specifies the theory, together with $a$.\\ \\
However, our aim being to address the problem of motion of two compact bodies in interaction, it will prove convenient to ``skeletonize" them, that is to replace $S_{\rm m}$ in (\ref{actionEMD}) by a phenomenological point particle action, $S_{\rm m}^{\rm pp}$.
Now, extending the treatment of Eardley in scalar-tensor theories \cite{Eardley} to incorporate the presence of the vector field, we shall consider the most generic covariant ansatz 
\begin{align}
S_{\rm m}^{\rm pp}[g_{\mu\nu},A_\mu,\varphi, \{x_A^\mu\}]=-\sum_A\int\! m_A(\varphi)\,ds_A+\sum_A q_A\int\! A_\mu\, dx^\mu_A\ ,\label{ansatz}
\end{align}
where $ds_A=\sqrt{-g_{\mu\nu} dx_A^\mu dx_A^\nu}$, $x_A^\mu[s_A]$ being the worldline of particle $A$, and where $m_A(\varphi)$ is a scalar function to be related to the skeletonized body $A$ later, which depends on the value of the scalar field evaluated at its location $x_A^\mu(s_A)$ (substracting divergent self contributions). In constrast, $q_A$ will be considered as a constant parameter. Indeed, as one easily sees from (\ref{ansatz}), the linear coupling of the worldlines to the vector field $A_\mu$ preserves the $U(1)$ gauge symmetry provided that the following current, $j^\mu$, is conserved:
\begin{equation}
\partial_\mu j^\mu=0\ ,\qquad \text{where}\qquad j^\mu(y)=\sum_A q_A\,\delta^{(3)}(\vec y-\vec{x}_A(t))\frac{dx_A^\mu}{dt}\ .
\end{equation} 
This in turn implies, as usual, that the charge $q_A$ of each body is conserved (provided that it remains separated from its companion), and cannot depend upon the variation of $\varphi(\vec x_A(t))$.\\

Note that our ansatz (\ref{ansatz}) does not depend on the gradients of the fields and, therefore, cannot take into account any finite size (e.g., tidal) effect, that we will neglect in the present paper; see, e.g., \cite{Damour:1998jk}. The functions $m_A(\varphi)$ must therefore be understood as depending on the homogeneous, adiabatically varying, scalar field environment of the body $A$, created by the faraway and slowly orbiting companion.\\

Finally, in the weakly self-gravitating limit, the bodies follow the geodesics of the Jordan metric, see (\ref{actionEMD}) and below. Their mass functions are hence reduced, in that limit, to $m_A(\varphi)=\tilde m_A \mathcal A(\varphi)$, where $\tilde m_A=cst$ is the Jordan-frame mass. In contrast, general relavitity is recovered when $m_A(\varphi)=cst$, together with $q_A=0$.

 \subsection{The post-Keplerian Lagrangian\label{sectionLagrangian1PK}}
 We now have in hands the necessary material to address the dynamics of compact binary systems in EMD theories: our starting point is the ``skeleton" action
 \begin{align}
 S[g_{\mu\nu},A_\mu,\varphi,\{x_A^\mu\}]&=\frac{1}{16\pi}\int d^4x \sqrt{-g}\bigg(\! R-2g^{\mu\nu}\partial_\mu\varphi\partial_\nu\varphi-e^{-2a\varphi}F^2\bigg)-\sum_A\int m_A(\varphi)\,ds_A+\sum_A q_A\int A_\mu\, dx^\mu_A\ ,
 \end{align}
with $ds_A=\sqrt{-g_{\mu\nu} dx_A^\mu dx_A^\nu}$, that depends on the fondamental parameter $a$, together with two mass functions $m_A(\varphi)$ and two constant charges $q_A$.
The field equations are now sourced by the effective particles:
\begin{subequations}
\begin{align}
&R_{\mu\nu}=2\partial_\mu\varphi\partial_\nu\varphi+2e^{-2a\varphi}\left(F_{\mu\alpha}F_\nu^{\ \alpha}
-\frac{1}{4}g_{\mu\nu}F^2\right)+8\pi\sum_A\left(T^A_{\mu\nu}-\frac{1}{2}g_{\mu\nu}T^A\right)\ ,\label{twoBody_fieldEqn_Einstein}\\
&\nabla_\nu\left(e^{-2a\varphi}F^{\mu\nu}\right)=4\pi\sum_A q_A\int\! ds_A\,\frac{\delta^{(4)}\left(y-x_A(s_A)\right)}{\sqrt{-g}}\,\frac{dx_A^\mu}{ds_A}\ ,\\
&\Box\,\varphi=-\frac{a}{2}e^{-2a\varphi}F^2+4\pi\sum_A\int\! ds_A\,\frac{dm_A}{d\varphi}\frac{\delta^{(4)}\left(y-x_A(s_A)\right)}{\sqrt{-g}}\ ,
\end{align}\label{twoBody_fieldEqn}%
\end{subequations}
where $\delta^{(4)}\left(x-y\right)$ is the 4-dimensional Dirac distribution, and where $T^A_{\mu\nu}$ is the stress-energy tensor associated to the skeletonized body $A$,
\begin{equation}
T_A^{\mu\nu}=\int\! ds_A\, m_A(\varphi)\frac{\delta^{(4)}\left(y-x_A(s_A)\right)}{\sqrt{-g}}\frac{dx_A^\mu}{ds_A}\frac{dx_A^\nu}{ds_A}\ .
\end{equation}

In this paper, we shall focus on the conservative part of the dynamics, neglecting all radiation reaction forces. Moreover, for bound orbits, it is convenient to implement relativistic corrections to the Keplerian dynamics in the weak field, slow orbital velocity limit.

\vfill\eject

\noindent In appendix \ref{Appendix_Lagr_1PK}, we derive the first post-Keplerian (1PK), $\mathcal O(m/R)\sim \mathcal O(V^2)$ corrections to the two-body Lagrangian (where $R$ is the distance separating the bodies, and $V$ denotes their relative orbital velocity), extending to EMD theories the standard methods presented in, e.g., \cite{Damour:1992we} in massless scalar-tensor theories (ST).
%\footnote{Note that PK scheme is based on the use of ``mass functions" $m_A(\varphi)$ which are well suited to encompass strong self-gravity effects of, say, neutron stars or black holes; the post-Newtonian aproximation is recovered for small compacities ($s=m/r<<1$).}
There, the field equations (\ref{twoBody_fieldEqn}) are solved perturbatively around the Minkowski metric $\eta_{\mu\nu}$ and the constant value  $\varphi_0$ of the background scalar field which, being associated to no gauge symmetry, see (\ref{EMDvacuumAction}) and below, cannot be set equal to zero, and is imposed by the cosmological environment of the binary system.\\ \\
At 1PK order, it is also useful to define the following body-dependent quantities
\begin{align}
&\alpha_A(\varphi)\equiv\frac{d\ln m_A}{d\varphi}\ ,\qquad \beta_A(\varphi)\equiv\frac{d\alpha_A}{d\varphi}\ ,\label{sensitivities}\\
\text{such that}\qquad &\ln m_A(\varphi)=\ln m_A^0+\alpha_A^0(\varphi-\varphi_0)+\frac{1}{2}\beta_A^0 (\varphi-\varphi_0)^2+\cdots\ ,\nonumber
\end{align}
where and from now on, a zero superscript indicates a quantity evaluated at infinity, $\varphi=\varphi_0$, as in $m_A^0=m_A(\varphi_0)$. Note that the parameter $\alpha_A^0$ measures the effective coupling between the scalar field and the skeletonized body $A$, and will play an essential role in the following.\\

With the definitions given above, the post-Keplerian Lagrangian reads, in harmonic coordinates $\partial_\mu(\sqrt{-g}g^{\mu\nu})=0$, and in the Lorenz gauge $\nabla_\mu A^\mu=0$ (see, again, appendix \ref{Appendix_Lagr_1PK} for details):
\begin{align}
L_{AB}&=-m_A^0-m_B^0+\frac{1}{2}m_A^0V_A^2+\frac{1}{2}m_B^0V_B^2+\frac{G_{AB}m_A^0 m_B^0}{R}\label{Lag1PK_PPN}\\
&+\frac{1}{8}m_A^0V_A^4+\frac{1}{8}m_B^0V_B^4+\frac{G_{AB}m_A^0 m_B^0}{R}\left[\frac{3}{2}(V_A^2+V_B^2)-\frac{7}{2}(V_A.V_B)-\frac{1}{2}(N.V_A)(N.V_B)+\bar \gamma_{AB}(\vec V_A-\vec V_B)^2\right]\nonumber\\
&-\frac{G_{AB}^2 m_A^0 m_B^0}{2R^2}\left[m_A^0(1+2\bar\beta_B)+m_B^0(1+2\bar\beta_A)\right]+\mathcal O(V^6)\ ,\nonumber
\end{align}
where $R\equiv\vert \vec x_A-\vec x_B\vert$, $\vec N\equiv(\vec x_A-\vec x_B)/R$, and $\vec V_A\equiv d\vec x_A/dt$.
We have also introduced the combinations
\begin{subequations}
\begin{align}
&G_{AB}\equiv1+\alpha_A^0\alpha_B^0-e_Ae_B\ ,\label{gAB}\\ 
&\bar\gamma_{AB}\equiv\frac{-4\,\alpha_A^0\alpha_B^0+3\,e_A e_B}{2(1+\alpha_A^0\alpha_B^0-e_Ae_B)}\ ,\label{PPNgamma}\\
&\bar\beta_A\equiv\frac{1}{2}\frac{\beta_A^0{\alpha_B^0}^2-2\,e_Ae_B(a\,\alpha_B^0-\alpha_A^0\alpha_B^0)+e_B^2(1+a\,\alpha_A^0-e_A^2)}{(1+\alpha_A^0\alpha_B^0-e_Ae_B)^2}\ ,\label{PPNbeta}
\end{align}\label{PPN}
\end{subequations}
and $A\leftrightarrow B$ counterpart, together with the convenient notations
\begin{equation}
e_A\equiv\frac{q_A}{m_A^0}e^{a\varphi_0}\quad\text{and}\ \quad e_B\equiv\frac{q_B}{m_B^0}e^{a\varphi_0}\ .\label{ratio}
\end{equation}

The expression (\ref{Lag1PK_PPN}-\ref{ratio}) of the post-Keplerian Lagrangian in EMD theories, which results from our generic ``skeleton" ansatz (\ref{ansatz}), is the first new technical result of this paper.\\

Remarkably, (\ref{Lag1PK_PPN}) has exactly the same structure than the 1PK Lagrangian in ST theories, see, e.g., \cite{Damour:1992we}. Indeed, the effects of the scalar \textit{and} vector fields have been entirely gathered in the four body-dependent combinations (\ref{PPN}), which generalize those introduced by Damour and Esposito-Far\`ese in ST theories \cite{Damour:1992we}, that we retreive when the charges vanish, $q_{A/B}=0$. The effective (dimensionless, since we set $G=1$) gravitational coupling $G_{AB}$ reflects the addition of the metric, scalar and (repulsive) vector interactions at the linear level, and is reduced to unity in the general relativity limit, $q_{A/B}=0$ and $\alpha^0_{A/B}=\beta^0_{A/B}=0$. The combinations $\bar\gamma_{AB}$ and $\bar\beta_{A/B}$ depend quadratically on the couplings to the scalar field $\alpha_{A/B}^0$ and to the vector field $e_{A/B}$, and vanish in the GR limit.\\ \\
 Finally, we note that our Lagrangian also encompasses Einstein-Maxwell's theory, which is retrieved when the bodies decouple from the scalar field, $m_{A/B}=cst$, i.e. $\alpha^0_{A/B}=\beta^0_{A/B}=0$.
 
 \vfill\eject

In this section, we proposed a generic ``skeleton" ansatz (\ref{ansatz}) that is well-suited to address the problem of motion of two arbitrary compact bodies in EMD theories (e.g., neutron stars, or black holes). At 1PK order, we showed that the resultant two-body Lagrangian (\ref{Lag1PK_PPN}) only differs from the ST one through the redefinition of $G_{AB}$, $\bar\gamma_{AB}$ and $\bar\beta_{A/B}$ given in (\ref{PPN}), which makes their associated two-body dynamics \textit{a priori} undistinguishable at this order.\\ \\
However, EMD theories deserve special attention
since, contrarily to ST theories \cite{Hawking:1972qk}, they predict the existence of ``hairy" black holes (i.e. black holes that differ from the general relativistic ones), that are encompassed in the generic approach developped above. Consequently, our aim in the next sections will be to specify our results for systems involving such ``hairy" black holes, that is, to compute explcitely the values for their post-Keplerian parameters $\alpha_A^0$, $\beta_A^0$ and $e_A$ entering our Lagrangian (\ref{Lag1PK_PPN}).

\section{Reducting a hairy black hole to a point particle\label{sectionSkeletonization}}

In the following, we will restrict our attention to a specific subclass of ``hairy" black holes with an electric charge, presented in subsection \ref{subsHairyBH}. In order to address their motion in presence of a companion, we shall, in a second step, ``skeletonize" them, that is, compute explicitly the effective ``mass function" $m_A(\varphi)$ and parameter $q_A$ that enter the effective point particle action describing them,
\begin{align}
S_{\rm m}^{\rm pp}(A)=-\int\! m_A(\varphi)\,ds_A+ q_A\int\! A_\mu\, dx^\mu_A\ ,
\end{align}
see (\ref{ansatz}) and the discussion below.\\

To this extent, we will ``zoom in" to the near-wordline region of particle $A$, which will take on the role of the black hole, and match the effective fields it generates to the real black hole solution. In particular, in keeping with neglecting all finite-size effects (see, again, section \ref{subsectionSkeleton}), the only \textit{gauge-invariant} influence of the faraway companion $B$ will be to impose a value for the adiabatically varying scalar environment $\varphi_\infty$ of our black hole.

\subsection{Electrically charged dilatonic black holes\label{subsHairyBH}}
Isolated black holes, that is, solutions of the vacuum field equations (\ref{vacuumEqs}), have been thoroughly studied in the literature. In particular, static, spherically symmetric black holes with an electric and (or) magnetic charge were introduced in \cite{Gibbons:1987ps} and \cite{Garfinkle:1990qj}, while their axisymmetric counterparts were found in \cite{Frolov:1987rj} and \cite{Horne:1992zy} in Kaluza-Klein theory (i.e. when $a=\sqrt{3}$ only).\\

In this paper, we shall restrict our attention to the class of electrically charged, non-spinning black hole solutions that read, in Just coordinates,
\begin{subequations}
\begin{align}
&ds^2=-\left(1-\frac{r_+}{r}\right)\left(1-\frac{r_-}{r}\right)^{\frac{1-a^2}{1+a^2}}dt^2+\left(1-\frac{r_+}{r}\right)^{-1}\left(1-\frac{r_-}{r}\right)^{-\frac{1-a^2}{1+a^2}}dr^2+r^2\left(1-\frac{r_-}{r}\right)^{\frac{2a^2}{1+a^2}}d\Omega^2\ ,\\
&A_t=-\frac{Q\,e^{2a\varphi_\infty}}{r}\ ,\quad A_i=0\ ,\\
&\varphi=\varphi_\infty+\frac{a}{1+a^2}\ln\left(1-\frac{r_-}{r}\right)\ ,
\end{align}\label{TNsolution}%
\end{subequations}
where $d\Omega^2=d\theta^2+\sin^2\theta\, d\phi^2$ and where $Q$ is not an independent integration constant but rather satisfies
\begin{equation}
Q^2=\frac{r_+ r_-}{1+a^2}\,e^{-2a\varphi_\infty}\ .\label{Q_relation}
\end{equation}
Note that $Q$ is the conserved $U(1)$ charge of the black hole, as one easily sees from a direct integration of (\ref{maxwellVacuum}).
Note also that we have ``gauged away" the asymptotic value of $A_\mu$ to zero for convenience.\\

 The solution (\ref{TNsolution}-\ref{Q_relation}) hence depends on three integration constants: the radius $r_+$ of the horizon, the location $r_-$ of the curvature singularity, and the asymptotic value $\varphi_\infty$ of the scalar field which, in the presence of the faraway companion, identifies to the local, adiabatically varying value of the scalar field it produces.\\

We note that the electric charge is crucial to induce our black hole with a scalar ``hair". Indeed, when $r_-=0$, (\ref{TNsolution}) is reduced to Schwarzschild's black hole. Another important limit is the scalar-vector decoupling, $a=0$, see (\ref{EMDvacuumAction}). In that case, $\varphi=\varphi_\infty$, $Q^2=r_+r_-$, and (\ref{TNsolution}) is reduced to Reissner-Nordstr\"om's black hole.\\

Finally, as part of the ``skeletonization" to come below, it will prove sufficient to expand the solution (\ref{TNsolution}) at infinity and in isotropic coordinates ($r=\tilde{r}+[r_++r_-]/2+\cdots$) as
\begin{subequations}
\begin{align}
&\tilde{g}_{\mu\nu}=\eta_{\mu\nu}+\delta_{\mu\nu}\left(\frac{r_++\frac{1-a^2}{1+a^2}\,r_-}{\tilde{r}}\right)+\mathcal{O}\left(\frac{1}{\tilde r^2}\right)\ ,\\
&A_t=-\frac{Q\,e^{2a\varphi_\infty}}{\tilde r}+\mathcal{O}\left(\frac{1}{\tilde r^2}\right)\ ,\\
&\varphi=\varphi_\infty-\frac{a}{1+a^2}\frac{r_-}{\tilde r}+\mathcal{O}\left(\frac{1}{\tilde r^2}\right)\ .
\end{align}\label{reel_asympt}%
\end{subequations}
Indeed, we shall see in the next subsection that the asymptotic expansion (\ref{reel_asympt}), which depends on the three integration constants $r_+$, $r_-$ and $\varphi_\infty$, encodes all the necessary information to skeletonize the black hole, i.e. to fix uniquely the function $m_A(\varphi)$ and constant $q_A$ of the effective particle that will take on the role of the black hole.

\subsection{The matching conditions\label{subsectionMatching}}

In the near-wordline region of the effective particle $A$, and at leading order in the large separation limit, the effective field equations (\ref{twoBody_fieldEqn}) are reduced to
\begin{subequations}
\begin{align}
&R_{\mu\nu}=2\partial_\mu\varphi\partial_\nu\varphi+2e^{-2a\varphi}\left(F_{\mu\alpha}F_\nu^{\ \alpha}
-\frac{1}{4}g_{\mu\nu}F^2\right)+8\pi\left(T^A_{\mu\nu}-\frac{1}{2}g_{\mu\nu}T^A\right)\ ,\\
&\nabla_\nu\left(e^{-2a\varphi}F^{\mu\nu}\right)=4\pi q_A\int\! ds_A\,\frac{\delta^{(4)}\left(y-x_A(s_A)\right)}{\sqrt{-g}}\,\frac{dx_A^\mu}{ds_A}\ ,\\
&\Box\,\varphi=-\frac{a}{2}e^{-2a\varphi}F^2+4\pi\int\! ds_A\,\frac{dm_A}{d\varphi}\frac{\delta^{(4)}\left(y-x_A(s_A)\right)}{\sqrt{-g}}\ ,
\end{align}\label{skelFieldEqn}
\end{subequations}
where we recall that $ds_A=\sqrt{-g_{\mu\nu}dx_A^\mu dx_A^\nu}$ and that
\begin{equation}
T_A^{\mu\nu}=\int\! ds_A\, m_A(\varphi)\frac{\delta^{(4)}\left(y-x_A(s_A)\right)}{\sqrt{-g}}\frac{dx_A^\mu}{ds_A}\frac{dx_A^\nu}{ds_A}\ .\label{skelStressEnergyTensor}
\end{equation}
Let us solve these field equations perturbatively around
Minkowski's metric $\eta_{\mu\nu}$, a vector field that can also be ``gauged away" to zero, and a scalar background, $\varphi_\infty$, which is the remaining influence of body $B$, as in (\ref{reel_asympt}). In harmonic coordinates (that identify to isotropic coordinates at this order),
$\partial_\mu(\sqrt{-\tilde g}\tilde g^{\mu\nu})=0$, and in the rest-frame of the particle $A$, i.e. setting $\tilde{\vec x}_A=\vec 0$, this yields, at linear order,
\begin{subequations}
\begin{align}
&\tilde{g}_{\mu\nu}=\eta_{\mu\nu}+\delta_{\mu\nu}\left(\frac{2m_A(\varphi_\infty)}{\tilde{r}}\right)+\mathcal O\left(\frac{1}{\tilde{r}^2}\right)\ ,\\
&A_t=-\frac{q_A\,e^{2a\varphi_\infty}}{\tilde{r}}+\mathcal O\left(\frac{1}{\tilde{r}^2}\right)\ ,\\
&\varphi=\varphi_\infty-\frac{1}{\tilde r}\frac{dm_A}{d\varphi}(\varphi_\infty)+\mathcal O\left(\frac{1}{\tilde{r}^2}\right)\ ,
\end{align}\label{skel_asympt}
\end{subequations}
which depends on the four effective parameters $q_A$, $\varphi_\infty$, $m_A(\varphi_\infty)$ and $m_A'(\varphi_\infty)$.\\

Now, by comparing (\ref{reel_asympt}) to (\ref{skel_asympt}), one obtains the following matching conditions:
\begin{subequations}
\begin{align}
&m_A(\varphi_\infty)=\frac{1}{2}\left(r_++\frac{1-a^2}{1+a^2}\,r_-\right)\ ,\label{matching1}\\
&q_A=Q\ ,\label{matching2}\\
&\frac{dm_A}{d\varphi}(\varphi_\infty)=\frac{a}{1+a^2}\,r_-\ ,\label{matching3}
\end{align}\label{matching}%
\end{subequations}
where we recall that the electric charge $Q$ satisfies
\begin{equation}
Q^2=\frac{r_+ r_-}{1+a^2}\,e^{-2a\varphi_\infty}\ ,\label{rappelQ}
\end{equation}
see (\ref{Q_relation}).
 Therefore, one ``skeletonizes" the real black hole (\ref{TNsolution}) as an effective point particle, provided that $q_A$, $m_A$ and its derivative evaluated at $\varphi_\infty$ satisfy the three matching conditions (\ref{matching}).\\ \\
 Note that as expected, $q_A$ identifies to the electric charge $Q$ of the black hole, see (\ref{matching2}), while $m_A(\varphi_\infty)$ coincides with its standard, say, ADM mass \cite{Arnowitt:1962hi}, $M_{\rm ADM}=\left(r_++\frac{1-a^2}{1+a^2}\,r_-\right)/2$, see (\ref{matching1}), which will therefore not be conserved when the scalar environment $\varphi_\infty$ of the black hole $A$, as created by its companion $B$, varies along its orbit.\\
 
 Moreover, the system (\ref{matching}) is integrable. Indeed, inverting (\ref{matching}) to substitute $m_A$, $dm_A/d\varphi$ and $q_A$ to $r_+$, $r_-$ and $Q$ in (\ref{rappelQ}), yields the first order differential equation to be satisfied by the function $m_A(\varphi)$:
\begin{equation}
\left(\frac{dm_A}{d\varphi}\right)\left(m_A(\varphi)-\frac{1-a^2}{2a}\frac{dm_A}{d\varphi}\right)=\frac{a}{2}\,q_A^2\,e^{2a\varphi}\ .
\label{eq_non_lineaire}
\end{equation}
For a given theory $a$, the solution of (\ref{eq_non_lineaire}) depends on the charge $q_A$ of the black hole, together with a unique integration constant, that we shall denote $\mu_A$.
The obtention of the matching conditions (\ref{matching}), together with the differential equation (\ref{eq_non_lineaire}) is the second new technical result of this paper, which shows that the dynamics of the black hole (\ref{TNsolution}) is described by two constant parameters only, $q_A$ and $\mu_A$.

\section{The sensitivity of a hairy black hole\label{section_Sensitivities}}

In this section, we solve the differential equation (\ref{eq_non_lineaire}), that is, we compute explicitely the function $m_A(\varphi)$ that describes the dynamics of our ``hairy" black hole (\ref{TNsolution}) through the point particle action
\begin{align}
S_{\rm m}^{\rm pp}(A)=-\int\! m_A(\varphi)\,ds_A+ q_A\int\! A_\mu\, dx^\mu_A\ .
\end{align}
This will, in a second step, allow us to come back to the two-body Lagrangian (\ref{Lag1PK_PPN}) obtained in section \ref{sectionLagrangian1PK}, and compute the values of the fundamental post-Keplerian parameters $\alpha_A^0$, $\beta_A^0$ and $e_A$ that enter it. In particular, we will find that the dynamics of our black hole might depart significantly from that of black holes in general relativity.

\subsection{The theory $a=1$}
In the particular case $a=1$, (\ref{eq_non_lineaire}) is easily integrated as
\begin{equation}
m_A(\varphi)=\sqrt{\mu_A^2+ q_A^2 \frac{e^{2\varphi}}{2}}\ ,\label{exemple_typique}
\end{equation}
where $\mu_A$ is a positive integration constant. We hence have an explicit expression for the function $m_A(\varphi)$, which characterizes our black hole by means of two constant parameters:\\ \\
(i) its electric charge $q_A=Q$ (see (\ref{matching2})), given in (\ref{rappelQ}), which is essential for our black hole to carry scalar ``hair". Indeed, when $q_A=0$, (\ref{exemple_typique}) is reduced to $m_A(\varphi)=\mu_A$ and the black hole decouples both from the vector and the scalar fields; \\ \\
(ii) the constant $\mu_A$, whose physical interpretation can be obtained thus: expressing it in terms of $r_+$, $r_-$, $\varphi_\infty$, using the matching conditions (\ref{matching}), yields
\begin{equation}
\mu_A^2=\frac{r_+(r_+-r_-)}{4}=\frac{A_H}{16\pi}\ .\label{relationMuAire}
\end{equation}
Therefore, $\mu_A^2$ is proportional to the area $A_H$ of the horizon of the black hole.\\ \\
In other words, the skeletonization describes a black hole that, when submitted to an adiabatic variation of its scalar field environment $\varphi_\infty$ (created by the slowly orbiting, faraway companion),
readjusts its equilibrium  configuration (that is, $r_+$ and $r_-$) in order for both its electric charge and horizon area to remain constant (while we recall that its ADM mass is not conserved, see (\ref{matching1}) and comments below).\\

\begin{figure}[!h]
\caption{The ``Fermi-Dirac" coupling $\alpha_A(\varphi_0)$ between the black hole $A$ and the scalar field, as a function of its cosmological environment $\varphi_0$, when $a=1$. The ``scalarization" is centered on $\varphi^{\rm crit}_A=\ln\left|\mu_A\sqrt{2}/q_A\right|$. Three distinct black holes are represented, $|\mu_A/q_A|=\{10,10^{3},10^{5}\}$, $q_A$ and $16\pi\mu_A^2$ being respectively their conserved charges and areas.\label{fig:sub2}}
   \includegraphics[width=0.5\linewidth]{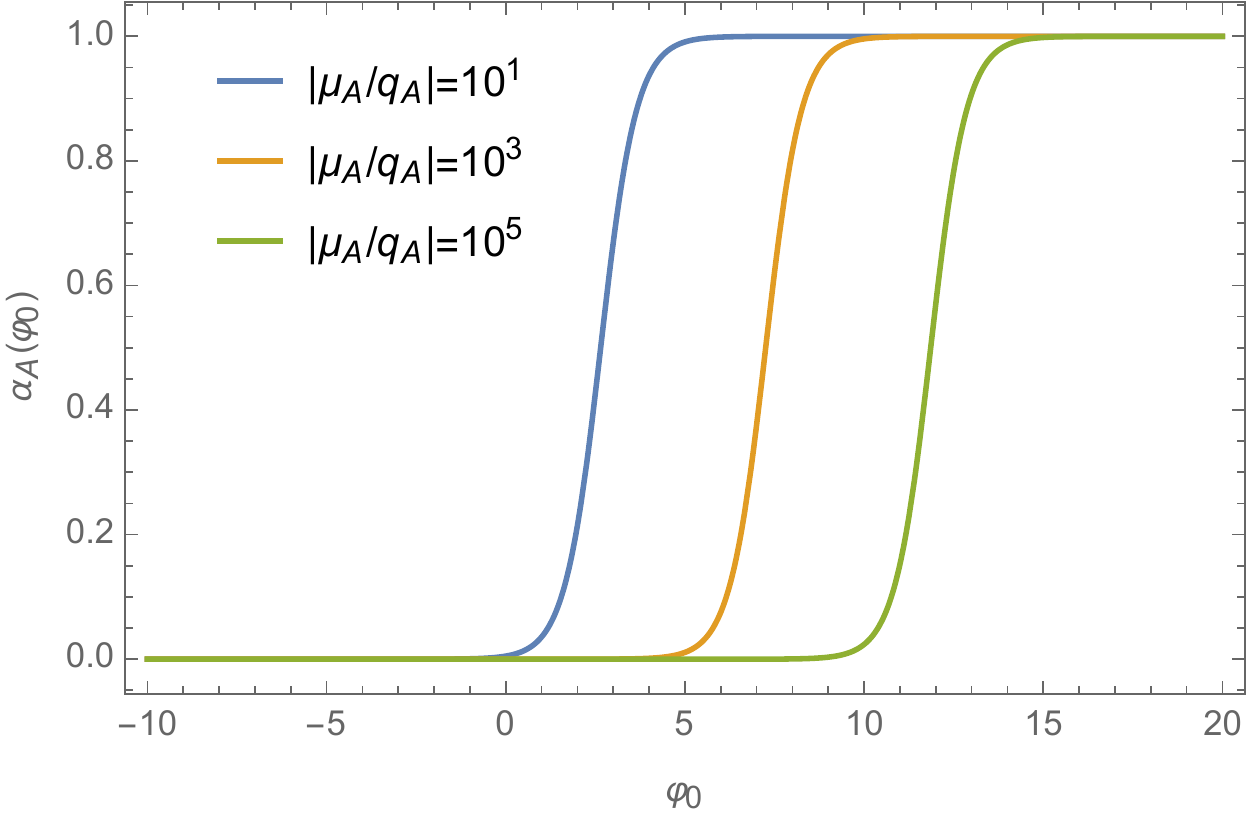}
\end{figure}

Moreover, given a specific black hole (i.e. specific values for $q_A$ and $\mu_A$), one can come back to the two-body problem, and compute the fundamental parameters that enter the 1PK Lagrangian (\ref{Lag1PK_PPN}), see (\ref{sensitivities}) and (\ref{ratio}),
\begin{equation}
\alpha_A^0=\frac{d\ln m_A}{d\varphi}(\varphi_0)\ ,\qquad \beta_A^0=\frac{d\alpha_A}{d\varphi}(\varphi_0)\ ,\qquad e_A=\frac{q_A}{m_A^0}e^{a\varphi_0}\ ,\label{rappelSensi}
\end{equation}  
in terms of $q_A$, $\mu_A$, and $\varphi_0$ which is the value of the scalar field far from the binary system, imposed by cosmology. Injecting the explicit mass function (\ref{exemple_typique}) in (\ref{rappelSensi}), we find
\begin{align}
&\alpha_A(\varphi_0)=\frac{1}{1+e^{2\left(\ln\left|\frac{\mu_A\sqrt{2}}{q_A}\right|-\varphi_0\right)}}\ ,\label{fermiDirac}\\ \nonumber\\
&\text{together with}\qquad \beta_A^0=2\alpha_A^0(1-\alpha_A^0)\qquad \text{and}\qquad (e_A)^2= 2\alpha_A^0\ .\label{betaA_eA}
\end{align}
\\
The simplicity of (\ref{fermiDirac}) is striking: the coupling $\alpha_A(\varphi_0)$ between the black hole and the scalar field is given by a ``Fermi-Dirac" distribution, as shown in figure \ref{fig:sub2}, which highlights the crucial influence of the scalar cosmological background $\varphi_0$ on the dynamics of a black hole. Indeed, when $\varphi_0$ is
increased through $\varphi^{\rm crit}_A=\ln\left|\mu_A\sqrt{2}/q_A\right|$,
the black hole $A$ transitions steeply between two extremal regimes:\\ \\
(a) a decoupled regime, $\alpha_A^0\rightarrow 0$, where it is moreover undistinguishable from the general relativistic Schwarzschild black hole, since $\alpha_A^0$ \textit{as well as} $\beta_A^0$ (and higher order derivatives of $\alpha_A$) and $e_A$ vanish, see (\ref{betaA_eA});\\ \\
(b) a regime where it is strongly coupled both to the scalar and to the vector fields, $\alpha_A^0\rightarrow 1$, together with $\beta_A^0\rightarrow 0$ and $(e_A)^2 \rightarrow 2$, which by definition, induces large deviations to the general relativistic two-body Lagrangian through the combinations $G_{AB}$, $\bar\gamma_{AB}$ and $\bar\beta_{A/B}$ that enter it, see (\ref{PPN}).\\ 

We note that the specific black hole considered (i.e., $q_A$ and $\mu_A$) only influences the location $\varphi^{\rm crit}_A=\ln\left|\mu_A\sqrt{2}/q_A\right|$ of the transition, while the values reached by $\alpha_A^0$, $\beta_A^0$ and $|e_A|$ in the extremal regimes (a) and (b) are \textit{universal}.

 From now on, this transition phenomenon will be referred to as ``scalarization", by analogy with the terminology introduced in the context of scalar-tensor theories to describe the spontaneous \cite{Damour:1992hw} and dynamical \cite{Palenzuela:2013hsa} scalarization of neutron stars (altough it must be noted that the phenomenon we highlight here is not a ``phase transition", and that the scalar coupling $\alpha_A^0$ transitions together with the vector coupling, $e_A$).

\subsection{Generic EMD theories\label{subsubTransition}}
\begin{figure}[!h]
   \caption{The coupling $\alpha_A(\varphi_0)$ between the black hole and the scalar field, as a function of its cosmological environment $\varphi_0$, for the theories $a\in \llbracket 0,10 \rrbracket$. The specific class of black holes such that $|\mu_A/q_A|=10^3$ is represented. When $a=0$, $\alpha_A=0$ and the black hole is Reissner-Nordst\"om's. For non-zero values of $a$, $\alpha_A(\varphi_0)$ becomes a function of $\varphi_0$ that transitions between $0$ and $a$, the steepness of the ``scalarization" increasing with $a$.\label{fig:sub1}}
   \includegraphics[width=0.55\linewidth]{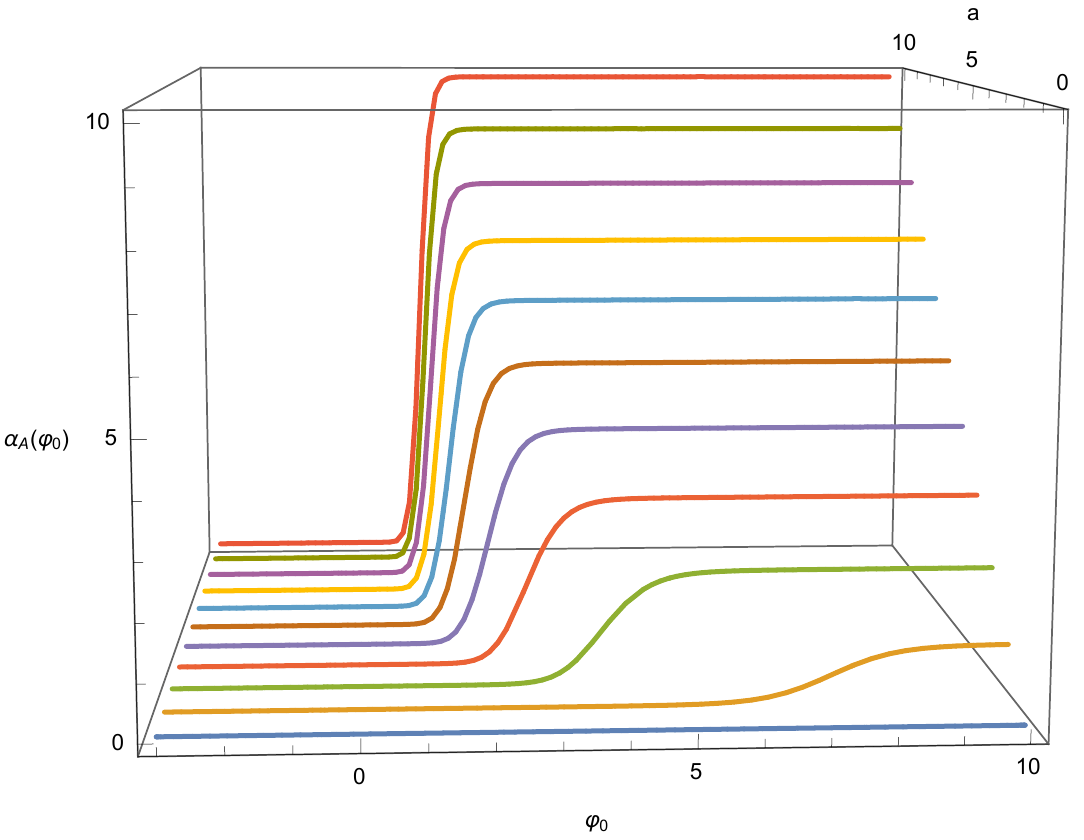}
\end{figure}

The remarkable features presented above are now easily transposed to generic EMD theories, that is, to an arbitrary value for the fundamental parameter $a$. First, one solves numerically the differential equation (\ref{eq_non_lineaire}), holding fixed the values for $q_A$ and for the integration constant $\mu_A$, in keeping with considering one specific black hole\footnote{Note that although we solve (\ref{eq_non_lineaire}) numerically, this differential equation admits an exact analytical solution in the form of a parametric equation on $m_A(\varphi)$, which is not enlightening to show here. It is however easy to find that the asymptotic behaviour of the solution is $m_A(\varphi)\underset{-\infty}{\sim} \mu_A$ and $m_A(\varphi)\underset{+\infty}{\sim} [q_A^2 e^{2a\varphi}/(1+a^2)]^{1/2}$, which is sufficient for the purpose of this paper.\\}. Then, one deduces the resultant coupling to the scalar field, $\alpha_A^0=d\ln m_A/d\varphi(\varphi_0)$, together with $\beta_A^0$ and $e_A$ through the relations\footnote{The obtention of (\ref{betaA_eA_generic}) is straightforward if one injects, by definition of $\alpha_A^0$ and $\beta_A^0$ (\ref{rappelSensi}), the expansion $m_A(\varphi)=m_A^0[1+\alpha_A^0(\varphi-\varphi_0)+\frac{1}{2}({\alpha_A^0}^2+\beta_A^0)(\varphi-\varphi_0)^2+\cdots]$ into the differential equation (\ref{eq_non_lineaire}), and solves it order by order to get
\begin{equation}
\alpha_A^0=\frac{a}{1-a^2}\left(1-\sqrt{1-e_A^2(1-a^2)}\right)\quad ,\qquad\beta_A^0=\frac{a^2e_A^2}{1-a^2} \bigg(1-a^2/\sqrt{\strut 1-e_A^2(1-a^2)}\bigg)\ ,\nonumber
\end{equation}
which is easily inverted to give (\ref{betaA_eA_generic}).\\}
\begin{equation}
\beta_A^0=\alpha_A^0\, (a-\alpha_A^0)\left[\frac{(1-a^2)\alpha_A^0-2a}{(1-a^2)\alpha_A^0-a}\right]\quad ,\qquad (e_A)^2=\alpha_A^0\left[\frac{2a-(1-a^2)\alpha_A^0}{a^2}\right]\ .\label{betaA_eA_generic}
\end{equation}
The resulting parameter $\alpha_A^0$ is shown in figure \ref{fig:sub1}, for $a\in \llbracket 0,10 \rrbracket$, setting $|\mu_A/q_A|=10^3$.
Again, for all $a$, the black hole transitions between two extremal regimes:\\ \\
(a) a decoupling, ``Schwarzschild-like" regime
$\alpha_A^0\rightarrow 0$, together with $\beta_A^0\rightarrow 0$ and $e_A\rightarrow 0$, see (\ref{betaA_eA_generic});\\ \\
(b) a \textit{universally} scalarized regime, $\alpha_A^0\rightarrow a$, with $\beta_A^0\rightarrow 0$ and $(e_A)^2\rightarrow 1+a^2$.\\ 

 Note that the curves $\alpha_A(\varphi_0)$ shown in figure \ref{fig:sub1} are slightly deformed ``Fermi-Dirac" distributions (with an exact identification when $a=1$ only, see discussion above), the steepness of the transition increasing with $a$. Note also that, for a fixed theory $a$, the value of $q_A$ and $\mu_A$ only influences the location $\varphi_A^{\rm crit}$ of the transition (which is shifted to greater $\varphi_0$ values when $|\mu_A/q_A|$ is increased) and the sign of $e_A$.\\
 
\vfill\eject

A numerical study of the dynamics of binary black holes in EMD theories has been performed recently in \cite{Hirschmann:2017psw}. There, the authors found that when $|\varphi_0|=10^{-10}$ and $a=\{1,10^3,3\times 10^3\}$, the influence of the scalar field on binary black holes can be neglected in understanding their dynamics, which in turn becomes hardly distinguishable from its general relativistic counterpart.
These features are consistent with the approach developped troughout the present paper: for example, when $|\mu_A/q_A|=10^3$, $|\varphi_0|=10^{-10}$ and $a=3\times 10^3$, we find vanishingly small values for $\alpha_A^0$, $\beta_A^0$ and $e_A$, see, again, figure \ref{fig:sub1}.\\ \\
However, our results above hint that this Schwarzschild-like behaviour might be considerably revised when the value of $\varphi_0$ is increased, as shown by the ``scalarization" phenomenon discussed above. Now, gravitational wave detectors such as LIGO-Virgo (or the forthcoming LISA detector) are designed to detect highly redshifted sources, where, indeed, the cosmological environment $\varphi_0$ might have had a non-negligible value (and that differs from today). In such situations, one could expect large deviations to the general relativistic black hole dynamics.\\

The obtention of the explicit mass function $m_A(\varphi)$ and charge $q_A$ that describe EMD black holes, together with the identification of the transition phenomenon regarding their coupling to the scalar and vector fields, or ``scalarization", is the third, and main result of this paper.

\subsection{Scalarized binary black holes\label{appendix_BBH_examples}}

As discussed above, in the presence of one (or two) black holes, the dynamics of binary systems might depart significantly from its general relativistic counterpart.
For the sake of illustration, let us consider a binary system composed of two ``scalarized" black holes , i.e. in the \textit{idealistic} limit when both are described by the regime (b) presented above, \textit{and} with electric charges $q_A$ and $q_B$ of the same sign:
\begin{equation}
 \alpha_{A/B}^0\rightarrow a\ ,\quad \beta_{A/B}^0\rightarrow 0\ ,\quad\text{and}\quad e_{A/B} \rightarrow \sqrt{1+a^2}\ ,\label{rappelCouplings}
\end{equation}
the value of the parameters ($q_A$, $\mu_A$) and ($q_B$, $\mu_B$) influencing only the location $\varphi_{A/B}^{\rm crit}$ of their transition, see the discussions above, and where an irrelevant sign was chosen in the definition of $e_{A/B}$.\\

 Injecting (\ref{rappelCouplings}) into the combinations $G_{AB}$, $\bar\gamma_{AB}$ and $\bar\beta_{A/B}$, given in (\ref{PPN}), that enter the post-Keplerian Lagrangian (\ref{Lag1PK_PPN}) presented in section \ref{sectionLagrangian}, yields
$G_{AB}\rightarrow 0$, $G_{AB}\,\bar\gamma_{AB}\rightarrow (3-a^2)/2$, and $G_{AB}^2\,\bar\beta_{A/B}\rightarrow 0$.\\ \\
 Therefore, all in all, the final two-body Lagrangian is reduced, at 1PK order, to
\begin{align}
L_{AB}&\rightarrow-m_A^0\sqrt{1-V_A^2}-m_B^0\sqrt{1-V_B^2}+\left(\frac{3-a^2}{2}\right)\frac{m_A^0 m_B^0}{R}(\vec V_A-\vec V_B)^2+\mathcal O(V^6)\ .\label{PapapetrouMajumdar}
\end{align}
The simplification (\ref{PapapetrouMajumdar}) is spectacular: among all the interaction terms present in the generic Lagrangian (\ref{Lag1PK_PPN}), only the $\bar\gamma_{AB}$-driven one remains. Moreover, this last term vanishes as well when the bodies are at relative rest ($\vec V_A-\vec V_B=\vec 0$). In other words, EMD binary black holes can transition to a universally ``scalarized" regime where their (attractive) metric, scalar and (repulsive) vector interactions compensate to allow for configurations at rest, at 1PK level at least.\\

This result is consistent with the Majumdar-Papapetrou spacetime\footnote{I am grateful to Thibault Damour for mentioning and sharing his expertise on the Majumdar-Papapetrou binary black hole solutions.}, which is a solution of the Einstein-Maxwell field equations that describes two \textit{extremal} Reissner-Nordstr\"om black holes at rest, see \cite{Majumdar:1947eu, Papaetrou:1947ib,Hartle:1972ya, Gibbons:1986cp}, and their extensions to EMD theories \cite{Gibbons:1987ps,Garfinkle:1990qj,Scherk:1979aj}. For a ``scalarizing" black hole, $e_A$ behaves as
\begin{equation}
(e_A)^2=\left(\frac{q_A}{m_A^0}\right)^2e^{2 a\varphi_0}\rightarrow 1+a^2\ ,\label{eq_extremalization}
\end{equation}
where $q_A=Q$ is the electric charge of the black hole, and $m_A^0=M_{\rm ADM}$ its ADM mass, see (\ref{matching}) and the discussion below, which generalizes the notion of an extremal black hole to generic EMD theories, $a\neq 0$: indeed, (\ref{eq_extremalization}) is equivalent to $r_-\rightarrow r_+$ (as proven using again the matching conditions (\ref{matching})).\\ \\

In other words, our ``scalarizing" black holes are in fact transitioning towards a ``quasi-extremal" regime (but never reach an exact extremal, ``naked singularity" configuration, since $\varphi_0$ is always finite).\\

Another striking consequence of (\ref{PapapetrouMajumdar}) is that in the context of Kaluza-Klein theory, i.e., $a=\sqrt{3}$, the last interaction term vanishes as well. Therefore, each component of the scalarized black hole binary behaves as a free particle at 1PK order. The generalization of this phenomenon to higher PK order remains to be investigated.

\subsection{General relativity and Einstein-Maxwell theory\label{subsection_RN_S}}
Let us conclude this section by mentioning that, as it should, the approach developped above is consistent with the well-known general relativity and Einstein-Maxwell theory limits.\\ \\
(i) Firstly, for any EMD theory, i.e. any value for $a$, but when the charge of the black hole vanishes, $q_A=0$, the three matching conditions (\ref{matching}) obtained in section \ref{subsectionMatching} are reduced to 
\begin{equation}
m_A=\frac{r_+}{2}=cst.\quad , \qquad q_A=Q=0\ .
\end{equation}
Therefore, the mass function $m_A$ identifies to the constant ``Schwarzschild" mass, our skeletonized black hole decouples both from the vector and scalar fields, and describes a Schwarzschild black hole, as expected from (\ref{TNsolution}) and the discussion below. This corresponds to the general relativity limit.\\ \\
(ii) Secondly, for any charge $q_A$, but in the scalar-vector decoupling limit $a=0$, see (\ref{EMDvacuumAction}), the matching conditions (\ref{matching}) become
\begin{equation}
m_A=\frac{r_++r_-}{2}=cst.\quad ,\qquad q_A=Q\ .
\end{equation}
This time, our effective particle describes a Reissner-Nordstr\"om black hole, as predicted in Einstein-Maxwell theory, see again discussion below (\ref{TNsolution}).

\section{Concluding remarks}

Reducing compact bodies to effective point particles has proven to be a very efficient tool to address analytically the many-body problem in general relativity and scalar-tensor theories. In this paper, we generalized this ``skeletonization" to Einstein-Maxwell-dilaton (EMD) theories, and showed that the corresponding most generic skeleton ansatz was the combination of two well-known point particle actions: that introduced by Eardley in scalar-tensor theories, which consists in endowing point particles with a specific ``mass function" $m_A(\varphi)$, and that of a charged particle in electrodynamics, endowed with a constant charge $q_A$.\\

More importantly, we computed, for the first time, the mass functions $m_A(\varphi)$ to be assigned to black holes, and shed light on two major properties regarding their dynamics: firstly, that it is encoded into two parameters (per black hole) only, $q_A$ and $\mu_A$, which are related to their charge and area\footnote{Note that we proved that $\mu_A$ is related to the area of the black hole for the specific theory $a=1$, see (\ref{relationMuAire}). Its generalization to arbitrary EMD theories will be the topic of \cite{CardenasJulie}.}; secondly, that, in keeping their charge and area fixed, black holes can undergo a new type of ``scalarization" phenomenon, leading to a transition between a regime where their dynamics is undistinguishable from that of Schwarzschild's black holes, and a ``quasi-extremal" regime where they strongly and universally couple to the scalar and vector fields (to within the sign of their electric charge). We note that the simplicity of our results, as illustrated by, e.g., figure \ref{fig:sub2}, is inherent to black holes, and contrasts with the rather involved study of neutron stars and their scalarization, see \cite{Damour:1992hw}, \cite{Barausse:2012da} or \cite{Palenzuela:2013hsa} (which, for example, depends on their equation of state, and on their coupling to the Jordan metric, $\tilde g_{\mu\nu}=\mathcal A^2(\varphi)g_{\mu\nu}$, see (\ref{actionEMD})).

\vfill\eject

As already discussed in section \ref{section_Sensitivities}, the ``scalarization" of a given black hole depends drastically on the value $\varphi_0$ of its cosmological environment, which, in turn, is expected to vary in the large range of redshifts that will be reachable with the gravitational wave detectors LIGO-Virgo (and forthcoming LISA); see, for example, \cite{Damour:1993id} for the cosmological evolution of $\varphi_0$ in scalar-tensor theories (to which EMD theories can be fairly supposed to reduce to on cosmological scales).
The fact that, depending on the redshift from which, say, a binary black hole system emitted, its dynamics can be undistinguishable from the general relativistic one, or, rather, significantly depart from it, supports the importance to investigate sources emitting from the broadest possible range of redshifts in the future.\\

In another paper \cite{Julie:2017pkb}, we extended the effective-one-body (EOB) approach (see, e.g., \cite{Buonanno:1998gg} and \cite{Damour:2016bks}) to massless scalar-tensor theories (ST); that is, we reduced the ST two-body dynamics, which is known at second post-Keplerian order \cite{Mirshekari:2013vb}, to the geodesic motion of a test particle in an effective static, spherically symmetric metric. In doing so, we implemented the impact of ST theories on the coalescence of compact binary systems as parametrized corrections to the best available general relativistic (5PN) EOB results. Now, in the present paper, we computed the two-body Lagrangian at 1PK level and showed that it only differs from the 1PK ST one in redefining the specific combinations $G_{AB}$, $\bar\gamma_{AB}$, and $\bar\beta_{A/B}$, see (\ref{PPN}) and discussions below. In consequence, the ST-EOB results, presented in \cite{Julie:2017pkb}, are trivially extended to EMD theories (including black holes). For example, the EMD correction to the orbital frequency of a compact binary system at the innermost stable circular orbit (ISCO), presented in \cite{Julie:2017pkb} figure 1 in ST theories, can be obtained replacing simply $G_{AB}$, $\bar\gamma_{AB}$, and $\bar\beta_{A/B}$ by their EMD values.\\

Finally, we note that, although the point particle action (\ref{ansatz}) from which we started, together with the matching conditions (\ref{matching}), have fixed uniquely the mass function $m_A(\varphi)$ and charge $q_A$ to be assigned to a specific EMD black hole, their validity remain to be justified.
Now, as discussed in section \ref{subsectionSkeleton}, our point particle action (\ref{ansatz}) does not depend on the gradients of the fields; this presupposes in turn that, for a well-separated binary system, each body is adiabatically readjusting its equilibrium configuration when submitted to the influence of its slowly orbiting, faraway companion.
 In \cite{CardenasJulie}, the thermodynamics of EMD black holes in equilibrium will provide a convenient framework to interpret and justify their skeletonization.
 
 \section*{Acknowledgments}
I am very grateful to Nathalie Deruelle for bringing EMD theories to my attention, for enlightening discussions and for her constant support.
I also acknowledge stimulating discussions with Marcela C\'ardenas.
Finally, I would like to thank Gilles Esposito-Far\`ese and Thibault Damour for very fruitful suggestions.

\appendix

\section{The two-body Lagrangian at 1PK order\label{Appendix_Lagr_1PK}}
In this appendix, we derive the two-body Lagrangian in EMD theories at post-Keplerian (1PK) order (\ref{Lag1PK_PPN}). The method presented below is an adaptation of the standard one given in \cite{Damour:1992we} in the scalar-tensor case.
For bound orbits, we shall implement the relativistic corrections to the Keplerian Lagrangian in the weak field, slow orbital velocity approximation, $\mathcal O(m/R)\sim \mathcal O(V^2)$, $R$ being the distance separating the bodies, and $V$ denoting their orbital velocity.\\

The first step is to solve the covariant field equations (\ref{twoBody_fieldEqn}) sourced by the two skeletonized bodies at 1PK order. To do so, let us rewrite them as
\begin{subequations}
\begin{align}
&R_{\mu\nu}=2\partial_\mu\varphi\partial_\nu\varphi+2e^{-2a\varphi}\left(F_{\mu\alpha}F_\nu^{\ \alpha}
-\frac{1}{4}g_{\mu\nu}F^2\right)+8\pi\sum_A \left(T^A_{\mu\nu}-\frac{1}{2}g_{\mu\nu}T^A\right)\ ,\label{EinsteinAppendix}\\
&\sqrt{-g}\,\nabla_\nu\left(e^{-2a\varphi}F^{\mu\nu}\right)=4\pi\sum_A q_A\,\delta^{(3)}\left(\vec{y}-\vec{x}_A(t)\right)\frac{d x_A^\mu}{dt}\ ,\label{MaxwellAppendix}\\
&\sqrt{-g}\,\Box\varphi=-\frac{a}{2}\sqrt{-g}\,e^{-2a\varphi}F^2+4\pi\sum_A \frac{ds_A}{dt}\frac{dm_A}{d\varphi}\delta^{(3)}\left(\vec{x}-\vec{x}_A(t)\right)\ ,\label{KGappendix}
\end{align}
\label{field_eqn_2body}
\end{subequations}
where we recall that $\Box\equiv \nabla^\mu\nabla_\mu$, where
\begin{equation}
T_A^{\mu\nu}= m_A(\varphi)\frac{\delta^{(3)}\left(\vec y-\vec x_A(t)\right)}{\sqrt{g g_{\alpha\beta}\frac{dx_A^\alpha}{dt}\frac{dx_A^\beta}{dt}}}\frac{dx_A^\mu}{dt}\frac{dx_A^\nu}{dt}
\end{equation}
is the stress-energy tensor of body $A$, and where $x^\mu_A=(t,\,\vec x_A)$ denotes its location.\\

In cartesian coordinates, the metric is now conveniently expanded around Minkowski's as
\begin{subequations}
\begin{align}
&g_{00}=-e^{-2U}+\mathcal O(V^6)\ ,\\
&g_{0i}=-4\,g_i+\mathcal O(V^5)\ ,\\
&g_{ij}=\delta_{ij}\,e^{2U}+\mathcal O(V^4)\ ,
\end{align}
\end{subequations}
(this ansatz allows to bypass unnecessary computational complications, see \cite{Blanchet:1989ki} and \cite{Damour:1990pi}) where, as we shall consistently check below, $U=\mathcal O(V^2)$ and $g_i=\mathcal O(V^3)$.\\
Similarly, the vector field is expanded around $A_\mu(r\rightarrow \infty)=0$, that is
\begin{equation}
A_\mu=\big(\delta A_t+\mathcal O(V^6)\ ,\ \delta A_i+\mathcal O(V^5)\big)\ ,
\end{equation}
where $\delta A_t=\mathcal O(V^2)$ and $\delta A_i=\mathcal O(V^3)$.\\
Finally, the scalar field is expanded around the local value of the cosmological scalar background, $\varphi_0$, as
\begin{equation}
\varphi=\varphi_0+\delta\varphi+\mathcal O(V^6)\ ,
\end{equation}
where $\delta\varphi=\mathcal O(V^2)$.\\

Therefore, the fonctions $m_A(\varphi)$ are expanded around $\varphi=\varphi_0$ at 1PK order, following the definitions (\ref{sensitivities}), as:
\begin{equation}
m_A(\varphi)=m_A^0\left[1+\alpha_A^0(\varphi-\varphi_0)+\frac{1}{2}({\alpha_A^0}^2+\beta_A^0)(\varphi-\varphi_0)^2+\mathcal O(V^6)\right]
\end{equation}
where we recall that a zero superscript indicates a quantity evaluated at $\varphi_0$, which is the value of the scalar field at spatial infinity.\\

In the harmonic gauge $\partial_\mu(\sqrt{-g}g^{\mu\nu})=0$, the $tt$ and $ti$ components of Einstein's field equations (\ref{EinsteinAppendix}) then read\footnote{Note that the harmonic condition $\partial_\mu(\sqrt{-g}g^{\mu\nu})=0$ implies $\partial_t U+\partial_i g_i=0$. Other useful intermediate results are $\sqrt{-g}=e^{2U}+\mathcal O(V^4)$, $R^{tt}=-\Box_\eta U +\mathcal O(V^6)$, $R^{ti}=-2\Delta g_i+\mathcal O(V^5)$, as well as $g_{\mu\nu}F^{t\mu}F^{t\nu}=(\partial_i A_t)(\partial_i A_t)+\mathcal O(V^6)$ and $F^2=-2(\partial_i A_t)(\partial_i A_t)+\mathcal O(V^6)$.}
\begin{subequations}
\begin{align}
&\Box_\eta U=-4\pi \sum_A m_A^0\left[1+\frac{3}{2}V_A^2-U+\alpha_A^0(\varphi-\varphi_0)\right]\delta^{(3)}(\vec x-\vec x_A(t))-(\partial_iA_t)(\partial_iA_t)e^{-2a\varphi_0}+\mathcal O(V^6)\ ,\\
&\Delta g_i=-4\pi\sum_A m_A^0 V_A^i \delta^{(3)}(\vec x-\vec x_A(t))+\mathcal O(V^5)\ ,
\end{align}
\end{subequations}
where $\Box_\eta\equiv\eta^{\mu\nu}\partial_\mu\partial_\mu$ denotes the flat Dalembertian. The Maxwell field equations (\ref{MaxwellAppendix}) read, in the Lorenz gauge $\nabla_\mu A^\mu=0$ (such that $\nabla_\nu F^{\mu\nu}=R^\mu_{\ \nu} A^\nu-\Box A^\mu$):
\begin{subequations}
\begin{align}
&\Box_\eta A_t=4\pi \sum_A q_A e^{2a\varphi_0}\bigg[1-2U+2a(\varphi-\varphi_0)\bigg] \delta^{(3)}(\vec x-\vec x_A(t))+2a(\partial_i A_t)(\partial_i \varphi)-2(\partial_i A_t)(\partial_i U)+\mathcal O(V^6)\ ,\\
&\Delta A_i =-4\pi\sum_A q_A e^{2a\varphi_0} V_A^i  \delta^{(3)}(\vec x-\vec x_A(t))+\mathcal O(V^5)\ .
\end{align}
\end{subequations}
Finally, the scalar field equation (\ref{KGappendix}) is
\begin{equation}
\Box_\eta \varphi=4\pi\sum_A m_A^0\alpha_A^0\left[1-\frac{1}{2}V_A^2-U+\left(\alpha_A^0+\frac{\beta_A^0}{\alpha_A^0}\right)(\varphi-\varphi_0)\right]\delta^{(3)}(\vec x-\vec x_A(t))+a(\partial_i A_t)(\partial_i A_t)e^{-2a\varphi_0}+\mathcal O(V^6)\ ,
\end{equation}
where $\Delta=\delta^{ij}\partial_i\partial_j$ denotes the flat Laplacian.\\

We now solve these equations, working iteratively. In particular, restricting ourselves to the conservative part of the dynamics, we shall use the half-retarded, half-advanced Green's function,
\begin{equation}
\Box_\eta G(x-x')=-4\pi\delta^{(3)}(\vec x-\vec x')\delta(t-t')
\end{equation}
where 
\begin{align}
G(x-x')&=\frac{1}{2}\left[\frac{\delta(t-t'-|\vec x-\vec x'|)}{|\vec x-\vec x'|}+\frac{\delta(t-t'+|\vec x-\vec x'|)}{|\vec x-\vec x'|}\right]\\
&=\delta(t-t')\left[\frac{1}{|\vec x-\vec x'|}+\frac{1}{2}\partial_t^2|\vec x-\vec x'|+\cdots\right]\ .
\end{align}
The right hand side of the above equations also contains extended sources that are easily integrated when using the identity \cite{Pati:2002ux}
\begin{align}
&f(\vec x,\, \vec y_1,\, \vec y_2)\equiv\frac{1}{2|\vec x-\vec y_1||\vec x-\vec y_2|}-\frac{1}{2|\vec y_1-\vec y_2|}\left(\frac{1}{|\vec x-\vec y_1|}+\frac{1}{|\vec x-\vec y_2|} \right)\quad \Rightarrow\quad \Delta f =\partial_i \frac{1}{|\vec x-\vec y_1|}\partial_i \frac{1}{|\vec x-\vec y_2|}\ ,
\end{align}
which is easily shown using Leibniz' rule.\\

Hence, all in all, the fields evaluated at any point $x^\mu=(t,\vec x)$ are found to be
\begin{subequations}
\begin{align}
&U(x)=\sum_A\frac{m_A^0}{\rho_A}\bigg[1+\frac{3}{2}V_A^2- \sum_{B\neq A} \frac{m_B^0}{R}(1+\alpha_A^0\alpha_B^0)\bigg]-e^{2a\varphi_0}\sum_{A,B}q_A q_B \,f\left(\vec x,\, \vec x_A(t),\, \vec x_B(t)\right)+\mathcal O(V^6)\ ,\label{fields_1PK_U}\\
& g_i(x)=\sum_A\frac{m_A^0}{|\vec x-\vec x_A(t)|} V_A^i+\mathcal O(V^5)\ ,
\end{align}
\end{subequations}
\begin{subequations}
\begin{align}
&A_t(x)=-e^{2a\varphi_0}\sum_A \frac{q_A}{\rho_A}\bigg[1-2\sum_{B\neq A}\frac{m_B^0}{R}(1+a\,\alpha_B^0)\bigg]+2 e^{2a\varphi_0}\sum_{A,B} q_A m_B^0 (1+a\,\alpha_B^0)f\left(\vec x,\, \vec x_A(t),\, \vec x_B(t)\right)+\mathcal O(V^6)\ ,\\
&A_i(x)=e^{2a\varphi_0}\sum_A\frac{q_A}{|\vec x-\vec x_A(t)|}V_A^i+\mathcal O(V^5)\ ,
\end{align}
\end{subequations}
and, finally,
\begin{align}
&\varphi(x)=\varphi_0-\sum_A \frac{m_A^0\alpha_A^0}{\rho_A}\bigg[1-\frac{1}{2}V_A^2-\sum_{B\neq A}\frac{m_B^0}{R}\left(1+\alpha_A^0\alpha_B^0-\frac{\beta_A^0\alpha_B^0}{\alpha_A^0}\right)\bigg]+a\, e^{2a\varphi_0}\sum_{A,B}q_A q_B f\left(\vec x,\, \vec x_A(t),\, \vec x_B(t)\right)+\mathcal O(V^6)\ ,\label{fields_1PK_scal}
\end{align}
where
\begin{align}
\frac{1}{\rho_A}&\equiv\frac{1}{|\vec x-\vec x_A(t)|}+\frac{1}{2}\partial_t^2|\vec x-\vec x_A(t)|\label{distance_propre}\\
&=\frac{1}{|\vec x-\vec x_A(t)|}\left[1+\frac{1}{2}\vec V_A^2-\frac{1}{2}(N_A.V_A)^2\right]+\frac{1}{2}(N_A.A_A)\ ,\nonumber
\end{align}
with $\vec N_A\equiv (\vec x_A-\vec x)/|\vec x_A-\vec x|$ and $\vec A_A\equiv d\vec V_A/dt$.\\

The obtention of the two-body Lagrangian is now straightforward. First, one writes the Lagrangian of, say, body $B$ when considered as a test particle in the fields generated by body $A$, and defined as ${S_{\rm m}^{\rm pp}}(B)\equiv\int\! dt\, L_B$, see, e.g., (\ref{ansatz}):
\begin{align}
L_B&=-m_B(\varphi)\,\frac{ds_B}{dt}+q_B\, A_\mu\, \frac{dx^\mu_B}{dt}\label{lagrangien_test_body}\\
&=-m_B(\varphi)\,\sqrt{e^{-2U}+8g_i V_B^i-e^{2U}V_B^2}+q_B (A_t+A_i V_B^i)+\mathcal O(V^6)\ ,\nonumber
\end{align}
setting formally $m_B^0=0$, $q_B=0$ and $\vec x=\vec x_B$ in (\ref{fields_1PK_U})-(\ref{fields_1PK_scal}). In particular, (\ref{distance_propre}) is easily rewritten as
\begin{equation}
\frac{1}{\rho_A}=\frac{1}{R}\left[1+\frac{1}{2}(V_A.V_B)-\frac{1}{2}(N_A.V_A)(N.V_B)\right]+\frac{1}{2}\frac{d}{dt}(N.V_A)\ ,\label{distance_propre_deriv}
\end{equation}
where $R=|\vec x_A-\vec x_B|$, $\vec N=(\vec x_A-\vec x_B)/R$, and where the last term is a total derivative that can be neglected in the Lagrangian (\ref{lagrangien_test_body}).

\vfill\eject

In a last step, one symmetrizes $L_B$ with respect to $A\leftrightarrow B$ to obtain the total two-body Lagrangian:
\begin{align}
L_{AB}&=-m_A^0-m_B^0+\frac{1}{2}m_A^0V_A^2+\frac{1}{2}m_B^0V_B^2+\frac{m_A^0 m_B^0}{R}(1+\alpha_A^0\alpha_B^0)-\frac{\tilde q_A\tilde q_B}{R}\label{Lagr1PK_brut}\\
&+\frac{1}{8}m_A^0V_A^4+\frac{1}{8}m_B^0V_B^4\nonumber\\
&+\frac{m_A^0 m_B^0}{R}\bigg[\left(\frac{V_AV_B}{2}\right)\left(-7+\alpha_A^0\alpha_B^0 \right)+\left(\frac{V_A^2+V_B^2}{2}\right)\left(3-\alpha_A^0\alpha_B^0\right)-\left(\frac{(N.V_A)(N.V_B)}{2}\right)\left(1+\alpha_A^0\alpha_B^0\right)\bigg]\nonumber\\
&+\frac{\tilde q_A\tilde q_B}{R}\bigg[\left(\frac{V_AV_B}{2}\right)+\left(\frac{(N.V_A)(N.V_B)}{2}\right)\bigg]\nonumber\\
&-\frac{m_A^0 m_B^0}{2R^2}\bigg[ m_A^0\bigg((1+\alpha_A^0\alpha_B^0)^2+\beta_B^0{\alpha_A^0}^2\bigg)+m_B^0\bigg((1+\alpha_A^0\alpha_B^0)^2+\beta_A^0{\alpha_B^0}^2\bigg)\bigg]\nonumber\\
&+\frac{\tilde q_A\tilde q_B}{R^2}\bigg[m_A^0(1+a\,\alpha_A^0)+m_B^0(1+a\,\alpha_B^0)\bigg]-\frac{\tilde q_B^2}{2R^2}\bigg[m_A^0(1+a\,\alpha_A^0)\bigg]-\frac{\tilde q_A^2}{2R^2}\bigg[m_B^0(1+a\,\alpha_B^0)\bigg]\nonumber\ ,
\end{align}
where we introduced the convenient notation $\tilde q_A \equiv q_A e^{a\varphi_0}$.
Finally, introducing $e_A\equiv \tilde q_A/m_A^0=(q_A/m_A^0)e^{a\varphi_0}$, (\ref{Lagr1PK_brut}) is straightforwardly rewritten as (\ref{Lag1PK_PPN}).

\bibliographystyle{unsrt}

\begin{thebibliography}{10}



\bibitem{Abbott:2016blz}
B.P. Abbott et~al.
\newblock {Observation of gravitational waves from a binary black hole merger},
\newblock {\em Phys. Rev. Lett. 116 no.6, 061102 (2016)}, [arXiv:1602.03837].

\bibitem{Abbott:2016nmj}
B.P. Abbott et~al.
\newblock {GW151226: Observation of gravitational waves from a 22-Solar-mass
  binary black hole coalescence},
\newblock {\em Phys. Rev. Lett. 116 no.24, 241103 (2016)}, [arXiv:1606.04855].

\bibitem{Abbott:2017vtc}
B.P. Abbott et~al.
\newblock {GW170104: Observation of a 50-Solar-mass binary black hole
  coalescence at redshift 0.2},
\newblock {\em Phys.Rev.Lett. 118 no.22, 221101 (2017)}, [arXiv:1706.01812].

\bibitem{Abbott:2017oio}
B.P. Abbott et~al.
\newblock {GW170814: A three-detector observation of gravitational waves from a
  binary black hole coalescence},
\newblock {\em Phys.Rev.Lett. 119 no.14, 141101 (2017)}, [arXiv:1709.09660].

\bibitem{TheLIGOScientific:2017qsa}
B.P. Abbott et~al.
\newblock {GW170817: Observation of Gravitational Waves from a Binary Neutron
  Star Inspiral},
\newblock {\em Phys.Rev.Lett. 119 no.16, 161101 (2017) }, [arXiv:1710.05832].

\bibitem{lesHouches}
T. Damour.
\newblock {Gravitational radiation and motion of compact bodies},
\newblock {\em Gravitational Radiation, ed. N.Deruelle and T.Piran, Centre de
  Physique des Houches, North-Holland (1983)}.

\bibitem{Blanchet:2013haa}
L. Blanchet.
\newblock {Gravitational radiation from post-Newtonian sources and inspiralling
  compact binaries},
\newblock {\em Living Rev.Rel. 17, 2 (2014)}, [arXiv:1310.1528].

\bibitem{Eardley}
D.M. Eardley.
\newblock {Observable effects of a scalar gravitational field in a binary
  pulsar},
\newblock {\em Astrophys. J. Lett. 196, L59 (1975)}.


\bibitem{Damour:1992we}
T. Damour and G. Esposito-Far\`ese.
\newblock {Tensor multiscalar theories of gravitation},
\newblock {\em Class. Quant. Grav. 9, 2093-2176 (1992)}.

\bibitem{Damour:1995kt}
T. Damour and G. Esposito-Far\`ese.
\newblock {Testing gravity to second post-Newtonian order: a field theory
  approach},
\newblock {\em Phys. Rev. D53, 5541-5578 (1996)}, [gr-qc/9506063].

\bibitem{Mirshekari:2013vb}
S. Mirshekari and C.M. Will.
\newblock {Compact binary systems in scalar-tensor gravity~: equations of motion
  to 2.5 post-Newtonian order},
\newblock {\em Phys. Rev. D87, 084070 (2013)}, [arXiv:1301.4680].


\bibitem{Damour:1992hw}
T. Damour and G. Esposito-Far\`ese.
\newblock {Nonperturbative strong field effects in tensor-scalar theories of
  gravitation},
\newblock {\em Phys. Rev. Lett. 70, 2220-2223 (1993)}.

\bibitem{Barausse:2012da}
E. Barausse, C. Palenzuela, M. Ponce, and L. Lehner.
\newblock {Neutron-star mergers in scalar-tensor theories of gravity},
\newblock {\em Phys. Rev. D87, 081506 (2013)}, [arXiv:1212.5053].

\bibitem{Palenzuela:2013hsa}
C. Palenzuela, E. Barausse, M. Ponce, and L. Lehner.
\newblock {Dynamical scalarization of neutron stars in scalar-tensor gravity
  theories},
\newblock {\em Phys. Rev. D89, 044024 (2014)}, [arXiv:1310.4481].

\bibitem{Julie:2017ucp}
F.-L. Juli\'e.
\newblock {Reducing the two-body problem in scalar-tensor theories to the
  motion of a test particle : a scalar-tensor effective-one-body approach, (2017)}. [arXiv:1709.09742]

\bibitem{Hawking:1972qk}
S.W. Hawking.
\newblock {Black holes in the Brans-Dicke theory of gravitation},
\newblock {\em Commun. Math. Phys. 25, 167-171 (1972)}.

\bibitem{Gibbons:1987ps}
G.W. Gibbons and K. Maeda.
\newblock {Black holes and membranes in higher dimensional theories with
  dilaton fields},
\newblock {\em Nucl.Phys. B298, 741-775 (1988)}.

\bibitem{Garfinkle:1990qj}
D. Garfinkle, G.T. Horowitz, and A. Strominger.
\newblock {Charged black holes in string theory},
\newblock {\em Phys.Rev. D43 , 3140 (1991)}.

\bibitem{Frolov:1987rj}
V.P. Frolov, A.I. Zelnikov, and U. Bleyer.
\newblock {Charged rotating black hole from five-dimensional point of view},
\newblock {\em Annalen Phys. 44, 371-377 (1987)}.

\bibitem{Horne:1992zy}
J.H. Horne and G.T. Horowitz.
\newblock {Rotating dilaton black holes},
\newblock {\em Phys.Rev. D46, 1340-1346 (1992)}, [hep-th/9203083].

\bibitem{Hirschmann:2017psw}
E.W. Hirschmann, L. Lehner, S.L. Liebling, and C. Palenzuela.
\newblock {Black hole dynamics in Einstein-Maxwell-dilaton theory, (2017)}. [arXiv:1706.09875].

\bibitem{Damour:1998jk}
T. Damour and G. Esposito-Far\`ese.
\newblock {Gravitational wave versus binary-pulsar tests of strong field
  gravity},
\newblock {\em Phys. Rev. D58, 042001 (1998)}, [gr-qc/9803031].



\bibitem{Arnowitt:1962hi}
R.L. Arnowitt, S. Deser, and C.W. Misner.
\newblock {The dynamics of general relativity},
\newblock {\em Gen.Rel.Grav. 40, 1997-2027 (2008)}, [gr-qc/0405109].

\bibitem{Majumdar:1947eu}
S.-D. Majumdar.
\newblock {A class of exact solutions of Einstein's field equations}.
\newblock {\em Phys.Rev. 72, 390-398 (1947)}.

\bibitem{Papaetrou:1947ib}
A.~Papapetrou.
\newblock {A Static solution of the equations of the gravitational field for an
  arbitrary charge distribution}.
\newblock {\em Proc.Roy.Irish Acad.(Sect.A) A51, 191-204 (1947)}.

\bibitem{Hartle:1972ya}
J.B. Hartle and S.W. Hawking.
\newblock {Solutions of the Einstein-Maxwell equations with many black holes},
\newblock {\em Commun.Math.Phys. 26, 87-101 (1972)}.

\bibitem{Gibbons:1986cp}
G.W. Gibbons and P.J. Ruback.
\newblock {The motion of extreme Reissner-Nordstrom black holes in the low
  velocity limit}.
\newblock {\em Phys.Rev.Lett. 57, 1492  (1986)}.

\bibitem{Scherk:1979aj}
J. Scherk.
\newblock {Antigravity: a crazy idea ?},
\newblock {\em Phys.Lett. 88B, 265-267 (1979)}.

\bibitem{CardenasJulie}
M. Cardenas, F.-L. Juli\'e, and N. Deruelle.
\newblock {Thermodynamics versus dynamics of hairy black holes (in
  preparation)}.

\bibitem{Damour:1993id}
T. Damour and K. Nordtvedt.
\newblock {Tensor-scalar cosmological models and their relaxation toward
  general relativity},
\newblock {\em Phys. Rev. D48, 3436-3450 (1993)}.


\bibitem{Julie:2017pkb}
F.-L. Juli\'e and N. Deruelle.
\newblock {Two-body problem in scalar-tensor theories as a deformation of
  general relativity~: an effective-one-body approach},
\newblock {\em Phys. Rev. D95, 124054 (2017)}, [arXiv:1703.05360].

\bibitem{Buonanno:1998gg}
A.~Buonanno and T.~Damour.
\newblock {Effective one-body approach to general relativistic two-body
  dynamics},
\newblock {\em Phys. Rev. D59, 084006 (1999)}, [gr-qc/9811091].

\bibitem{Damour:2016bks}
T. Damour and A. Nagar.
\newblock {The effective-one-body approach to the general relativistic two body
  problem},
\newblock {\em Lect. Notes Phys., 905, 273-312 (2016)}.

\bibitem{Blanchet:1989ki}
L.~Blanchet and T.~Damour.
\newblock {Postnewtonian generation of gravitational waves}.
\newblock {\em Ann.Inst.H.Poincare Phys.Theor. 50, 377-408 (1989)}.

\bibitem{Damour:1990pi}
T. Damour, M. Soffel, and C.M. Xu.
\newblock {General relativistic celestial mechanics. 1. Method and definition
  of reference systems},
\newblock {\em Phys.Rev. D43, 3273-3307 (1991)}.

\bibitem{Pati:2002ux}
M.E. Pati and C.M. Will.
\newblock {PostNewtonian gravitational radiation and equations of motion via
  direct integration of the relaxed Einstein equations. 2. Two-body equations
  of motion to second postNewtonian order, and radiation reaction to 3.5
  postNewtonian order}.
\newblock {\em Phys.Rev. D65, 104008 (2002)}.

\end{thebibliography}

\end{document}